\documentclass[aps,prb,twocolumn,showpacs]{revtex4-1}
\usepackage{graphicx}
\usepackage{graphics}
\usepackage{amsmath}
\usepackage{amsfonts}
\usepackage{amssymb}
\usepackage{epstopdf}
\usepackage{makeidx}
\usepackage{epsfig}
\usepackage{bm}
\usepackage{color}
\usepackage{xcolor}
\usepackage[unicode=true, bookmarks=false,breaklinks=false,pdfborder={0 0 1},backref=false,colorlinks=false]{hyperref}
\usepackage{makeidx}
\hypersetup{colorlinks=true,citecolor=blue,linkcolor=blue,filecolor=blue,urlcolor=blue}
\newcommand{\be}   {\begin{equation}}
\newcommand{\ee}   {\end{equation}}
\newcommand{\ba}   {\begin{eqnarray}}
\newcommand{\ea}   {\end{eqnarray}}

\begin{document}

\title{Weiss oscillations in graphene with a modulated height profile}

\author{Rhonald Burgos}
\affiliation{Instituto de F\'{\i}sica,
 Universidade Federal Fluminense, 24210-346 Niter\'oi RJ, Brazil}
\author{Caio Lewenkopf}
\affiliation{Instituto de F\'{\i}sica,
 Universidade Federal Fluminense, 24210-346 Niter\'oi RJ, Brazil}
 
\date{\today}

\begin{abstract}
We study the electronic transport properties of a monolayer graphene with a one-dimensional 
modulated height profile caused, for instance, by substrate ondulations. 
We show that the combined effect of the resulting strain fields induce modulated scalar and 
vector potentials that give rise to Weiss oscillations in the magnetoconductivity. 
We also find that similar effects can be obtained by applying a parallel magnetic field to 
the graphene-substrate interface. 
The parameters of an experimental set-up for a physical realization of these findings  
in graphene systems are discussed.
\end{abstract}


\maketitle

\section{Introduction}
\label{sec:intro}

The magnetoresistivity of a two-dimensional electron gas (2DEG) subjected to a periodic 
potential varying in one direction shows very strong oscillations periodic in inverse magnetic 
field \cite{Weiss1989}. 
This remarkable effect, called Weiss oscillations, is observed in the magnetoresistivity parallel 
to the grating direction of the periodic potential and is negligible on the transversal and longitudinal 
directions. 
The effect was quantitatively in terms of the semiclassical electronic velocity obtained from the 
quantum mechanical analysis of the changes in the local band structure due to the modulated 
potential \cite{Winkler1989,Gerhardts1989,Vasilopoulos1989,Peeters1992}. 
The Weiss oscillations can also be understood using a classical approach that 
associates their periodicity with the commensurability of the cyclotron motion and the grating, 
which modifies the root-mean-square of the drift velocity of the guiding center \cite{Beenakker1989}.
This gives origin to oscillations of the magnetoresistivity with period $2R_c/\lambda$, where 
$R_c$ is the cyclotron radius and $\lambda$ is the period of the grating. This nice and 
intuitive picture is corroborated by the solution of the Boltzmann equation, assuming both
an isotropic \cite{Beenakker1989} and anisotropic \cite{Mirlin1998} disorder scattering processes. 

Several theoretical works studied Weiss oscillations in graphene systems. Using the quantum 
mechanical approach \cite{Winkler1989,Gerhardts1989,Vasilopoulos1989,Peeters1992}, oscillations 
in the magnetoconductivity were calculated for the cases of monolayer graphene sheet modulated 
magnetic~\cite{Tahir2008} and electric field~\cite{Matulis2007,Tahir2007,Tahir2007PRB}. 
The theory of Weiss oscillations was also extended to bilayer graphene~\cite{Peeters2012}.
These studies put in evidence the similarities and the differences between Weiss oscillations in 
graphene and 2DEG systems.
One of the conclusions is that one expects the effect to be more robust against temperature in 
graphene, due to its unique spectral properties. Unfortunately, there is no experimental report of 
Weiss oscillations in graphene so far. 

The main goal of this paper is to propose a set-up that allows to experimentally observe the effect.
Assuming a given modulated profile height varying along a single-direction we explore two mechanisms 
that give rise to a periodic potential, namely, strain and/or an in-plane external magnetic applied on the 
graphene sheet. 

Strain modifies the interatomic distances and, hence, the electronic structure of the material. 
Combining an effective microscopic model for the low-energy properties of electrons in graphene 
with the theory of elasticity, it has been shown 
\cite{Suzuura2002,Manes2007,Guinea2008,CastroNeto2009,Vozmediano10,Masir2013}
that the effects due to strain fields can be accounted for by a pseudo electric and pseudo magnetic 
fields, that are incorporated to the effective graphene Hamiltonian as a diagonal scalar and vector potentials, 
respectively. Recent papers have shown that these pseudo fields
give measurable contributions for transport properties\cite{Couto2014,Burgos2015}.

A magnetic field applied parallel to the modulated grephene sheet can also generate an effective 
periodic vector potential as long as $\lambda$ is much larger that the height profile amplitude, 
as discussed in Refs.~\onlinecite{Lundeberg2010,Burgos2015}.
Experimentally, modulated profile heights have been reported in suspended membranes \cite{Bao2009} 
and nanoripples \cite{Lee2013,Bai2014}. Another possibility is to lithographically produce trenches, 
defining a profile height on a given substrate. After deposition, the graphene sheet acquires a similar shape.

This paper is organized as follows. 
Section~\ref{sec:model} begins with a brief review of the effective theory of the low energy dynamics 
of electrons in graphene under a uniform perpendicular magnetic field. 
We discuss the modulated pseudomagnetic and pseudo electric fields due to strain in Section~\ref{sec:strain}. 
The expression for the modulated parallel magnetic field is obtained in Section~\ref{sec:inplane}.
In Section~\ref{sec:weissconductivity} we present analytical closed expressions for the 
Weiss oscillations due to modulated  pseudo electric and pseudo magnetic fields 
In Section~\ref{sec:results} we present our main results, discuss the validity range of the theory, 
establishing bounds to guide an optimal choice of the experimental set-up parameters to study the 
effect.  
Finally we present our conclusions in Section~\ref{sec:conclusions}.

\section{Theoretical background}
\label{sec:model}

Our model Hamiltonian reads
\be
H= H_0 + H',
\ee
 where $H_0$ accounts for the dynamics of low-energy electrons in graphene 
 monolayers under  a uniform external magnetic field and $H'$ is the effective
 Hamiltonian due to the modulated deformation of the graphene sheet.\\
 
In the presence of an external applied magnetic field, the effective Hamiltonian for 
low energy electrons in graphene reads  \cite{CastroNeto2009,Goerbig2011}
\begin{equation}
\label{eq:Dirac}
H_0=v_F \bm \sigma \cdot \left( {\bm p} + e {\bm A}_{\rm ext}\right),
\end{equation}
where $v_F\approx 10^{6}$m/s is the Fermi velocity and 
$\bm \sigma=(\sigma_x, \sigma_y)$ 
are Pauli matrices in the lattice subspace \cite{CastroNeto2009}. 

For a uniform magnetic field perpendicular to the graphene plane, $\bm B_{\rm ext}=B_\perp \hat{{\bf z}}$, 
the vector potential can be written in the Landau gauge 
\begin{equation}
\label{Landaugauge}
\bm A_{\rm ext}=B_\perp[(1-\alpha)y \hat{\bm x}+\alpha x \hat{\bm y}].
\end{equation}
We postpone the discussion of the most convenient choice of $\alpha$ to the next section.
 
In what follows we obtain the effective perturbation Hamiltonian $H'$ that describes the
effects of strain due to a periodic out-of-plane deformation of the graphene sheet given by
\begin{equation}
\label{highprofile}
h(x)=h_0 \cos (2\pi x/\lambda),
\end{equation}
where $\lambda$ is the modulation period and $h_0$ is the profile height amplitude.

\subsection{Strain induced magnetic and electric fields}
\label{sec:strain}

Strain modifies the graphene inter-atomic distances and changes its electronic 
properties. It has been shown \cite{Suzuura2002,Manes2007,Midtvedt2016} that 
strain effects in the electronic dynamics can be accounted for by introducing a vector 
gauge potential and a scalar potential in the effective Hamiltonian given by 
Eq.~\eqref{eq:Dirac}. 

\begin{figure}[h]
\centering \includegraphics[width=0.97\columnwidth]{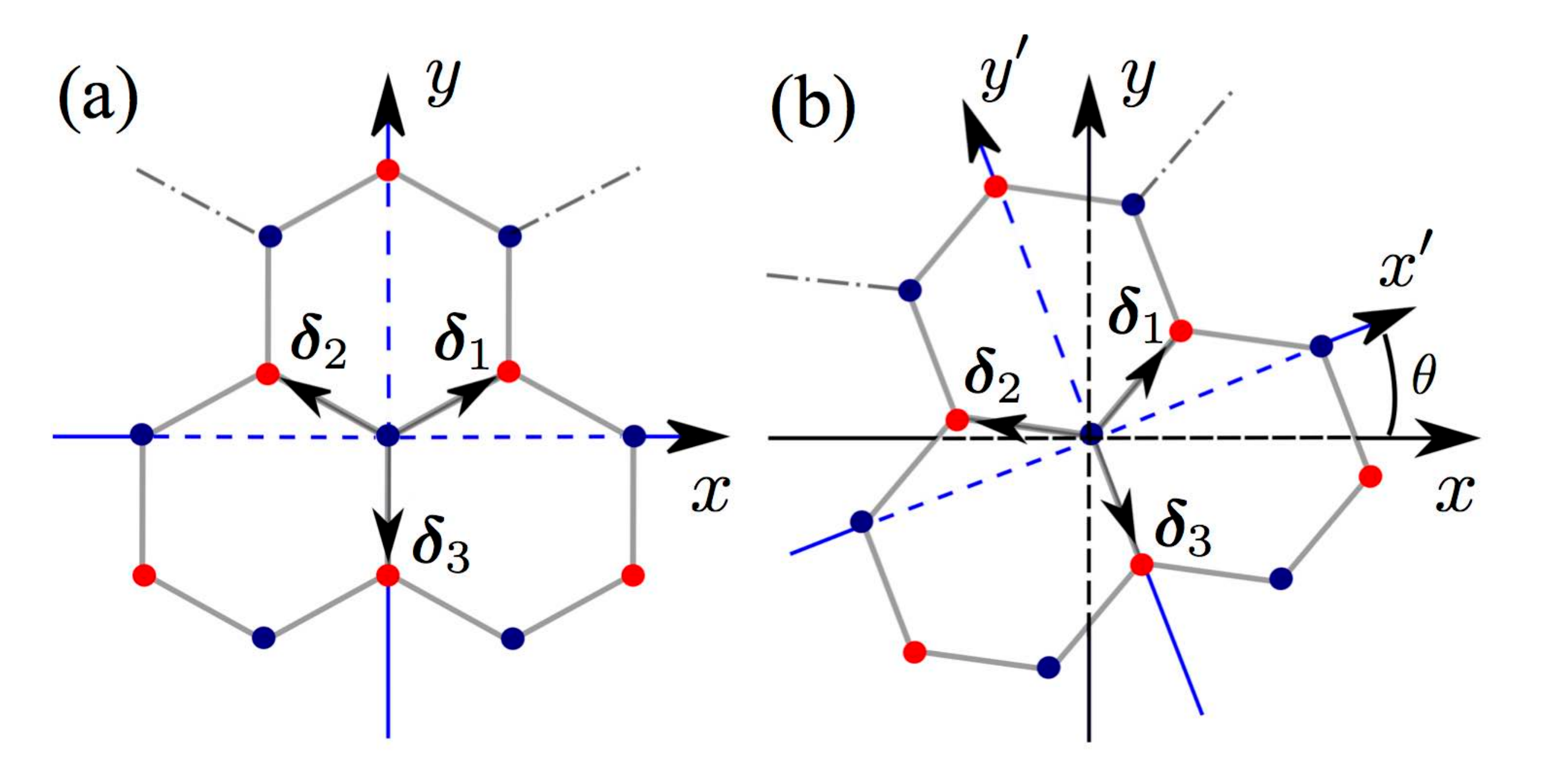}
\caption{(Color online) Honeycomb lattice orientation: (a) zigzag crystallographic 
orientation along the $x$-axis and (b) arbitrary orientation.}
\label{fig:lattice_orientation}
\end{figure}

By taking the long wavelength limit of the graphene tight-binding Hamiltonian the strain 
contribution to the system Hamiltonian, up to linear order in the deformations, can be 
cast as \cite{Suzuura2002,Manes2007,deJuan2013} 
\begin{equation}
\label{eq:H'_K1}
H'=t \sum_{n=1}^3
i\frac{\bm \sigma \cdot \bm \delta_n}{a}\sigma_z
\left(\frac{\beta_{\rm G}}{a^2}
\bm \delta_n ^T\cdot \bm u \cdot \bm \delta_n
\right),
\end{equation}
where ${\bm \delta}_n$ are the nearest neighbor vectors (see Fig.~\ref{fig:lattice_orientation}), 
$a\approx 1.42$ \AA ~is the carbon-carbon distance, 
$t\approx 2.7$ eV is the nearest neighbor  $\pi$-orbitals hopping matrix element,
and $\beta_{\rm G}=-\partial \log t/\partial \log a \approx 2-3.37$ \cite{Guinea2008,Pereira2009}
is the Gr\"uneisen parameter, a dimensionless material dependent parameter that characterizes 
the coupling between the Dirac electrons and the lattice deformations, and $\bm u$ is the strain 
tensor.

We express the components of the strain tensor in the ``macroscopic" $xy$ 
coordinate system. The lattice sites are more conveniently assigned by
``intrinsic" $x'y'$ coordinates oriented along the high symmetry crystallographic 
directions of the graphene lattice, as shown in Fig.~\ref{fig:lattice_orientation}a. 

In the intrinsic coordinate system the nearest neighbor vectors read \cite{CastroNeto2009}
\be
\bm \delta_1 = \frac{a}{2}
\left(\!\begin{array}{c} \sqrt{3}  \\ 1 \end{array} \!\right),\;
\bm \delta_2 =\frac{a}{2}
\left(\!\begin{array}{c}  -\sqrt{3}  \\  1 \end{array}\!\right),\;
\bm \delta_3 = a
\left(\! \begin{array}{c}  0  \\- 1 \end{array}\! \right).
\ee

For the case where $xy$ and $x'y'$ coincide, by inserting the relations for 
${\bm \delta}$ into Eq.~\eqref{eq:H'_K1}, the strain contribution to the system 
Hamiltonian becomes 
\be
H'= ev_F {\bm \sigma}\cdot {\bm A},
\ee
with
\begin{equation}
{\bf A} = \frac{ \hbar \beta_{\rm G} \kappa }{2ae} 
\left( \begin{array}{c}
u_{xx} - u_{yy} \\
 - 2u_{xy}
\end{array}\right),
\end{equation}
where $\kappa\approx 2/3$ stands for the correction of the graphene effective vector potential 
due to the violation of the Cauchy-Born rule in lattices with a basis \cite{Suzuura2002,Midtvedt2016}.

Let us now consider the more realistic case where the graphene zigzag crystal orientation 
forms an angle $\theta$ with the ``macroscopic" $x$-axis, see Fig.~\ref{fig:lattice_orientation}b. 
Accordingly, the nearest neighbor vectors ${\bm \delta}_n$ are rotated by $\theta$, namely
\be
\label{eq:delta_rotated}
 {\bm \delta}_n(\theta) = {\bf R}(\theta) \cdot {\bm \delta}_n,
\ee
where $ {\bf R}(\theta)$ is the rotation matrix in two-dimensions.
Inserting the rotated nearest neighbor vectors ${\bm \delta}_n(\theta)$ in 
Eq.~\eqref{eq:H'_K1} we obtain
\begin{equation}
{\bf A}(\theta) = \frac{ \hbar \beta_{\rm G} \kappa }{2ae} 
\left( \begin{array}{c}
(u_{xx} - u_{yy}) \cos 3\theta + 2 u_{xy} \sin3\theta \\
(u_{xx} - u_{yy}) \sin 3\theta - 2 u_{xy} \cos3\theta
\end{array}\right)
\end{equation}
or, in a more compact form, 
\begin{equation}
{\bf A} (\theta) = {\bf R}(3\theta) \cdot {\bf A}.
\end{equation}
We note that, with few exceptions \cite{Falko2012,Naumis,Verbiest2015}, the literature 
addresses only the perfect aligned case of $\theta=0$.

In addition to the pseudo vector potential, strain also induces a scalar potential 
\cite{Suzuura2002,vonOppen2009, Midtvedt2016} given by
\begin{equation}
\label{scalarpseudopotential}
V(\bm r)=g[u_{xx}(\bm r)+u_{yy}(\bm r)] I,
\end{equation}
where $I$ is the identity matrix in sublattice space and $g\approx 4$ eV 
\cite{Guinea2010}. 

The strain tensor components $u_{ij}(\bm r)$ read \cite{Suzuura2002,Manes2007,Guinea2008}
\begin{align}
u_{xx}({\bm r}) &=\frac{\partial u_x(\bm r)}{\partial x}+\frac{1}{2}\!\left[\frac{\partial h({\bm r})}{\partial x} \right]^2,
\nonumber\\
u_{yy}({\bm r}) &= \frac{\partial u_y(\bm r)}{\partial y}+\frac{1}{2}\!\left[\frac{\partial h({\bm r})}{\partial y}\right]^2,
\nonumber\\
u_{xy}({\bm r}) & = \frac{1}{2}\left[\frac{\partial u_x(\bm r)}{\partial y}+\frac{\partial u_y(\bm r)}{\partial x} \right]+\frac{1}{2}\frac{\partial h({\bm r})}{\partial x}\frac{\partial h({\bm r})}{\partial y}.
\end{align}
The in-plane displacement vector field ${\bm u} (\bm r)$ can be obtained, for instance, by minimizing the 
elastic energy \cite{Wehling2008,Guinea2008} for a given $h(\bm r)$, following the prescription 
proposed in Ref.~\onlinecite{Guinea2008}. 

For one-dimensional periodic modulations, such as 
$h(x)$ defined by Eq.~\eqref{highprofile}, the minimization of the elastic energy leads to a relaxed 
configuration where the strain tensor components become negligibly small \cite{Wehling2008}. 
Such analysis does not account for the fact that, in general, the graphene sheet is pinned to the substrate 
at random positions \cite{Ishigami2007} that introduce non trivial constraints on the in-plane displacements. 
In this paper we consider quenched ripples, setting $u_x(\bm r)=u_y(\bm r)=0$. We stress that this 
assumption gives an upper bound of the strain field and to the corresponding vector gauge potential.

The strain tensor corresponding to the out-of-plane deformation profile $h(x)$ described by 
Eq.~\eqref{highprofile} reads
\begin{eqnarray}
u_{xx}(x) &= &2\pi^2\left(\frac{h_0}{\lambda}\right)^2\sin^2(2\pi x/\lambda), 
\nonumber\\
u_{yy}(x) & = & u_{xy}(x)=0.
\end{eqnarray}

Hence, the pseudo vector potential reads 
\begin{equation}
{\bf A}(\theta) = A_0 \sin^2(2\pi x/\lambda) \left( \begin{array}{c}
\cos 3\theta  \\ \sin 3\theta 
\end{array}\right)
\end{equation}
where
\be
A_0 = \frac{\hbar \beta_{\rm G} \pi^2 \kappa}{ae} \left(\frac{h_0}{\lambda}\right)^2.
\ee
The pseudo scalar potential is given by
\begin{eqnarray}
\label{scalarpotential}
V(x)=V_0\sin^2(2\pi x/\lambda),
\end{eqnarray}
with 
\be
V_0=2 g\pi^2 \left(\frac{h_0}{\lambda}\right)^2.
\ee

We finish this section recalling that theoretical studies  \cite{Gibertini2012,Falko2012}
show that the scalar potential is dramatically screened by the carriers in the graphene flake.
Screening modifies the coupling $g$ as $g \rightarrow g/\epsilon(\bm q, \omega\rightarrow 0)$,
where 
$1/\epsilon(\bm q, \omega)=1+v(\bm q)\Pi^R(\bm q,\omega)$ with $\epsilon(\bm q, \omega)$ 
the dynamical dielectric function, $v(\bm q)=2\pi e^2/\epsilon_0|\bm q|$ the Coulomb interaction 
with $\epsilon_0$ being the substrate material dependent dielectric constant, and $\Pi^R(\bm q,\omega)$ 
is the retarded density-density correlation function. Within the random phase approximation (RPA) the 
dielectric function can be expressed as
$
\epsilon(\bm q, \omega) =1-u(\bm q)\chi^{0}(\bm q,\omega),
$
where $\chi^{0}(\bm q,\omega)$ is the pair bubble diagram. The static dielectric function 
reads \cite{DasSarma2007}
\be
\epsilon(\bm q,0) = 
\left\{ \begin{array}{ll} 
1 + v(\bm q)\rho(E_F), & \ q\leq 2k_F,\\
1 + v(\bm q)\rho(E_F)\Bigg[1-\frac{1}{2}\sqrt{1-\left(\frac{2k_F}{q}\right)^2} \\
\;\;\;\;\;\;\;\;\;\;\;
-\frac{q}{4k_F}\arcsin\left(\frac{2k_F}{q}\right)+
\frac{\pi q}{8k_F}\Bigg], & \,\,q> 2k_F.
\end{array} \right.
\ee
with $\rho(E_F)$ the density of states at the Fermi energy. We have checked that 
$\epsilon(\bm q, 0)\agt 40$ for graphene deposited on silicon dioxide, suggesting that
screening can strongly quench the pseudoelectric field. Based on this reasoning, 
it has been conjectured  \cite{Gibertini2012} that the random ripples, ubiquitous 
in deposited exfoliated graphene, are the cause of the charge puddles observed in 
these systems. Scanning tunneling microscopy (STM) experiments on graphene 
on SiO$_2$ \cite{Deshpande2009,Zhang2009} do not find evidences of spacial 
correlations between ripples and charge puddles and, thus, fail to support this 
picture.

In what follows we assume that screening is absent, corresponding to an upper bound
of the scalar potential. We address again the screening issue in Sec.~\ref{sec:results},
where we discuss an experimental set-up to measure Weiss oscillations.

\subsection{External in-plane magnetic field}
\label{sec:inplane}

A modulated magnetic field can also be realized by applying an external magnetic field parallel 
to a grated patterned graphene sheet. The external magnetic field has a component perpendicular 
to the graphene surface profile given by \cite{Burgos2015}
\begin{align}
	B' ({\bm r}) = 
	-{\bm B}_\parallel \cdot \hat{\bm n}({\bm r}).
\end{align}
The normal vector to the surface $z= h({\bm r})$ is 
\begin{align}
\hat{\bm n}({\bm r}) 
= \frac{1}{ \sqrt{ 1 + 
(\partial h/\partial x)^2 + (\partial h/\partial y)^2} } 
\left(\begin{array}{c}
\partial h ({\bm r})/\partial x \\ \partial h ({\bm r})/\partial y \\ -1 \end{array}\right).
\end{align}
Since $h_0 \ll \lambda$, we write
\be
\label{normalvector2}
\hat{\bm n}({\bm r}) \approx (
 \partial h({\bm r})/\partial x , \partial h({\bm r})/\partial y , -1 )^T.
 \ee
Hence, the effective local perpendicular magnetic field reads
 \be
 \label{externalmagneticfield}
 B_{\rm ext} ({\bm r}) =  -{\bm B}_\parallel \cdot {\bm \nabla} h({\bm r}),
 \ee 
that  for $\bm B_{\parallel}=B_{\parallel}\hat{\bm x}$ is expressed in a convenient gauge, 
by the vector potential
\begin{equation}
\label{eq:BparA}
A_x({\bm r}) = 0 \quad \mbox{and} \quad  A_y({\bm r}) = -A_{\|} \cos (2\pi x/\lambda)\,.
\end{equation}
with $A_{\|}=B_{\|}h_0$. The perturbation term is given by
\be
V_{\rm ext}(\bm{r}) = 
v_F e \sigma_y A_y({\bm r}) = -v_F e A_{\|} \cos (2\pi x/\lambda) \sigma_y.
\ee

\section{Weiss oscillations in graphene}
\label{sec:weissconductivity}

In this section we briefly review the calculations of the Weiss oscillations for modulated magnetic 
\cite{Tahir2008} and electric \cite{Matulis2007} fields, adapting the results to the vector and scalar 
fields  obtained in the previous section. 

We study the corrections to the conductivity caused by the modulated strain within the regime 
where the latter corresponds to a small perturbation of the electronic spectrum. In this case, one
can obtain an analytical expression for the Weiss conductivity oscillations following the approach 
put forward in Refs.~\onlinecite{Winkler1989,Gerhardts1989,Vasilopoulos1989}.

The scalar potential of Eq.~\eqref{scalarpotential} breaks the translational invariance along the 
$x$-axis. Hence, it is convenient to solve the unperturbed Hamiltonian $H_0$ in the Landau 
gauge with $\alpha=1$. Hence, the Schr\"odinger equation $H_0 \Psi(\bm r)=E\Psi(\bm r)$ has 
eigenvalues \cite{McClure1956,CastroNeto2009,Goerbig2011}
\begin{equation}
\label{eq:LL}
E_n={\rm sgn}(n)\,\hbar \omega_0\sqrt{2|n|},
\end{equation}
with $\omega_0 = v_F/l_B$ and
\be
{\rm sgn}(n) = \left\{ \begin{array}{ll} 
1 & \,\,n>0,\\
0 & \,\,n=0,\\
-1 & \,\,n<0.
\end{array} \right.
\ee
The corresponding eigenfunctions are \cite{Zheng2002}
\begin{equation}
\label{eigenstates}
\Psi_{n,k_y}(\bm r)=\frac{C_n}{\sqrt{L_yl_B}}e^{ik_y y}
\begin{pmatrix}
   -i{\rm sgn}(n)  \Phi_{|n|-1}(\frac{x-x_0}{l_B})  \\
      \Phi_{|n|   }(\frac{x-x_0}{l_B}) 
\end{pmatrix},
\end{equation}
where $l_B=\sqrt{\hbar/eB}\approx (26 \,{\rm nm})/\sqrt{B ({\rm T})}$ is the magnetic length, 
$x_0=l_B^2k_y$ gives the center of the wave function, 
\be
C_n = \left\{ \begin{array}{ll} 
1 & n=0,\\
1/\sqrt{2} & n\neq 0,
\end{array} \right.
\ee
and
\begin{equation}
\label{ }
 \Phi_{n}(x) =\frac{e^{-x^2/2}}{\sqrt{2^nn!\sqrt{\pi}}}H_n(x),
\end{equation}
where $H_n(x)$ are Hermite polynomials.

Starting from the Kubo formula for the conductivity, it has been shown \cite{Peeters1992}
that the main contribution to the Weiss oscillations comes from the diagonal diffusive
conductivity, that in the quasielastic scattering regime can be written as 
\cite{Charbonneau1982}
\begin{equation}
\label{eq:conductivity}
\Delta\sigma_{yy}=
g_vg_s
\frac{e^2}{L_xL_y}
\sum_{\zeta}
\left.\left(-\frac{\partial f}{\partial \varepsilon} \right)\right|_{\varepsilon=E_\zeta} 
\tau(E_{\zeta})
v_{\zeta,y}v_{\zeta,y},
\end{equation}
where $g_v$ and $g_s$ stand for valley and spin degeneracy (for graphene $g_vg_s=4$), 
$\zeta=(n,k_y)$ are the quantum numbers of the single-particle electronic states, 
$L_x$ and $L_y$ are the dimensions of the graphene layer,  $f(E_{\zeta})$ is the Fermi-Dirac 
distribution function, $\tau(E_{\zeta})$ is the electron relaxation time, and $v^{\zeta}_y$ is the 
electron velocity given by the semiclassical relation
\begin{equation}
\label{eq:v_semiclassic}
v_{\zeta,y}=\frac{1}{\hbar}\frac{\partial }{\partial k_y}E_{n,k_y},
\end{equation}
with $E_{n,k_y}$ calculated in first order perturbation theory as
\begin{equation}
\label{perturbation}
E_{n,k_y}=E_n+  \langle n,k_y | H' | n,k_y \rangle.
\end{equation}
Note that this correction lifts the degeneracy of Landau levels.
The dc diffusive conductivity is then obtained by explicitly summing over the quantum 
numbers, namely
\begin{equation}
\label{eq:sum}
\sum_\zeta \big[\cdots\big] = \frac{L_y}{2\pi} \int^{L_x/l_B^2}_0 dk_y \sum_{n=0}^\infty \big[\cdots\big].
\end{equation}

\subsection{Modulated scalar potential}
\label{scalarfield}

Let us now present the theory for Weiss oscillations for graphene monolayers in a modulated 
electric field \cite{Matulis2007}. We highlight the main results that are relevant to our 
analysis, deferring the details of the derivation to the original literature \cite{Matulis2007}.

For the scalar potential, the expectation value of the velocity operator 
$v_{\zeta,y}^{\rm s} = \hbar^{-1}\partial \langle n,k_y| V| n, k_y\rangle/\partial k_y$ reads
\begin{align}
\label{eq:v_a}
v_{\zeta,y}^{\rm s} &=\frac{2\pi V_0 l_B^2}{\hbar \lambda}
e^{-u/2} \nonumber \\
&\times
[{\rm sgn}^2(n)L_{|n|-1}(u)+L_{|n|}(u)]\sin\left(4\pi \frac{x_0}{\lambda}\right),
\end{align}
where $u=8 \pi^2 (l_B/\lambda)^2$ and $L_n(u)$ is a Laguerre polynomial. 

Inserting the above expression in Eq.~\eqref{eq:conductivity} and using \eqref{eq:sum}, 
$\Delta\sigma_{yy}$ reads
\be
\Delta\sigma^{\rm s}_{yy}
\approx
\frac{e^2}{h} \frac{V^2_0 \beta \tau }{ \hbar } F^{\rm s}(u, \beta, E_F)
\ee
with
\begin{align}
\label{Sumscalarcontribution}
F^{\rm s} =&
\frac{ue^{-u}}{\beta} \sum^{\infty}_{n=-\infty}
\left.\left(-\frac{\partial f}{\partial \varepsilon} \right)\right|_{\varepsilon=E_n} 
\!\!  \nonumber \\
&\times
[(1-\delta_{0,n})L_{|n|-1}(u)+L_{|n|}(u)]^2 
\end{align}
where $\beta = 1/k_B T$. We assume that $\Delta E_{n,k_y}=|E_{n,ky}-E_n|$ is smaller 
than the Landau level spacing and take $\tau=\tau(E_F)$.

Equation~\eqref{eq:conductivity} indicates that $\Delta \sigma_{yy}$ is dominated by the 
Landau levels with energies close to the Fermi energy.
In the limit where many Landau levels are either filled (for $ E_F>0 $) or empty (for $E_F<0$), 
it is possible \cite{Matulis2007} to obtain an analytical expression for $\Delta \sigma_{yy}^{\rm s}$. 
First, one uses the $n\gg 1$ asymptotic expression for the Laguerre polynomials \cite{Szego1955}
\be
\label{asymptotic}
e^{-u/2}L_n(u) \longrightarrow \frac{1}{\pi^{1/2} (nu)^{1/4}} \cos(2\sqrt{nu} - \pi/4),
\ee
that is very accurate as long as $u \alt n$. For $n \alt u \alt 4n$ the Laguerre polynomial is still 
an oscillatory function, but shows significant deviations from the expression given by 
Eq.~\eqref{asymptotic} \cite{Temme1990}. 
For $u \agt 4n$ the $L_n(u)$ is monotonic. For the Landau levels $n$ close to the Fermi energy,  
$u \agt 4n$ is translated to $k_F\lambda \alt 2\pi$, where the electrons become insensitive to 
the modulated potential, a regime that is hardly relevant for the realistic physical 
parameters, as discussed in the next section.

Next, one takes the continuum limit
\be 
\sum_{n=-\infty}^\infty \big[\cdots\big] \approx
\left(\frac{l_B}{v_F \hbar}\right)^2 \int_{-\infty}^{\infty} dE E \big[\cdots\big].
\ee

After some algebra we write
\be
\label{scalarintegrated}
F^{\rm s} =\frac{1}{\pi^2} \frac{T}{T_{\rm W} }\cos^2\!\left(\frac{2\pi}{k_F\lambda} \right)
\left[1 + S\!\left(\frac{T}{T_{\rm W}} \right) \sin\!\left(8\pi \frac{k_F l^2_B}{ \lambda}\right)\right],
\ee 
where 
\begin{equation}
\label{damp}
S(x)=\frac{x}{\sinh(x)} 
\quad \text{and} \quad
k_{\rm B}T_{\rm W}= \frac{\hbar \omega_0}{8\pi^2}\frac{\lambda}{l_B}.
\end{equation}

\subsection{Modulated vector potential}
\label{vectorfield}

Let us now turn our attention to the conductivity corrections due to magnetic modulations. 

In the case of the pseudo vector potential caused by strain, the expectation value of 
the velocity operator is 
\begin{align}
\label{eq:vpseudomagnetic}
v^{\rm v}_{\zeta,y}&= A_{0}
\frac{l_B v_Fe}{\hbar} {\sqrt{2}} \cos\!\left(\frac{4\pi}{\lambda}x_0\right)
\\ \nonumber
&\times 
e^{-u/2} {\rm sgn}(n) \sqrt{|n|+1}
\left[L_{|n|+1}(u)-L_{|n|}(u)\right]
\sin(3\theta),
\end{align}
where $u=8 \pi^2(l_B/\lambda)^2$ and $L_n(u)$ is a Laguerre polynomial. 
Inserting the results of Eqs.~\eqref{eq:vpseudomagnetic} and \eqref{vinplane} in 
Eq.~\eqref{eq:conductivity} 
and using \eqref{eq:sum}, the $\Delta\sigma^{\rm v}_{yy}$ reads
\be
\label{conductivityintrinsicmagneticfield}
\Delta\sigma^{\rm v}_{yy}
\approx
\frac{e^2}{h} \frac{(v_F e A_0)^2 \tau \beta}{\hbar} 
\sin^2(3\theta)\,
F^{\rm v}(u, \beta, E_F)
\ee
with
\begin{align}
\label{magneticcontributionnumeric}
F^{\rm v} =&
\frac{4e^{-u}}{\beta} \sum^{\infty}_{n=-\infty}\!\!
\left.\left(-\frac{\partial f}{\partial \varepsilon} \right)\right|_{\varepsilon=E_n} 
\!\!  (1-\delta_{0,n}) (|n|+1)
\nonumber \\
&\times \left[L_{|n|}(u)-L_{|n|+1}(u)\right]^2 .
\end{align}
Following the steps described in the scalar potential case, we obtain
\begin{align}
\label{Functionmagneticmodulation}
F^{\rm v} = & \frac{1}{4\pi^4}(\lambda k_F)^2 \frac{T}{T_{\rm W}}  
\sin^2\!\left(\frac{2\pi}{k_F\lambda} \right) 
\nonumber \\
& \times \left[1 - S\!\left(\frac{T}{T_{\rm W}} \right) \sin\left(8\pi \frac{k_F l^2_B }{\lambda} \right)
\right], 
\end{align} 
where $S(x)$ and $T_{\rm W}$ are defined in Eq.~\eqref{damp}.

For the case of an external in-plane magnetic field, the expectation of the velocity operator is 
\begin{align}
\label{vinplane}
v^{\|}_{\zeta,y} =& A_{\|} \frac{l_B v_Fe}{\hbar} \sqrt{2} \cos\left(\frac{2\pi}{\lambda} x_0\right)
\\ \nonumber 
& \times 
e^{-u/2} {\rm sgn}(n) \sqrt{|n|+1} \left[L_{|n|+1}(u)-L_{|n|}(u)\right],
\end{align}
with $u=2\pi^2 (l_B/\lambda)^2$. Note that the functional dependence of $v^{\|}_{\zeta,y}$ is 
very similar to the one of Eq.~\eqref{eq:vpseudomagnetic}, except for the periodicity which 
differs by a factor 2.
This observation allows us to readily write the conductivity correction for the case 
of an external modulated magnetic field as
\be
\label{conductivityExternalmagneticfield}
\Delta\sigma^{\rm \|}_{yy}
\approx
\frac{e^2}{h} \frac{(v_F e A_{\|})^2 \tau \beta}{\hbar} 
F^{\|}(u, \beta, E_F)
\ee
where $F^{\|}(u, \beta, E_F)$ is obtained by taking $\lambda \rightarrow 2\lambda$ in 
the expression for $F^{\rm v}(u, \beta, E_F)$.

\section{Results and discussion}
\label{sec:results}

In this section we discuss the validity range of our results and propose bounds for the parameter 
range of a set-up to realize the Weiss oscillations in graphene systems. We analyze separately 
the cases of Weiss oscillations caused by strain and those due to a parallel magnetic field.
We discuss their combined effect in a realistic experimental setup. 

We note that the obtained expressions for the Weiss oscillations are consistent with the 
semiclassical guiding-center-drift resonant picture due to Beenakker \cite{Beenakker1989}, 
that predicts
\begin{equation}
\label{Beenakkerequation}
\frac{\Delta \rho_{xx}}{\rho_{xx}}\propto
\cos^2\left(2\pi \frac{R_c}{\lambda^*}-\frac{\pi}{4} \right),
\end{equation}
where $R_c$ is the cyclotron radius and $\lambda^*$ the periodicity of the modulated potential. 
In the semiclassical regime of $n\gg1$, one can safely neglect {\it zitterbewegung} effects 
\cite{Cserti2006,Schliemann2008} and write $R_c=\hbar k_F/eB=\sqrt{2n}l_B$, 
in line with recent cyclotron orbits imaging observations \cite{Bhandari2016}. 
By recalling that $\lambda^*=\lambda/2$, one immediately identifies that the periodicity of the Weiss 
oscillations of Eq.~\eqref{Beenakkerequation} coincides with the expressions presented 
in the previous section.
This observation, so far overlooked in the graphene literature, suggests that the classical 
commensurability orbit resonance picture still holds in Dirac-like materials, as graphene. 

Let us now address the main assumption of the analysis presented in Sec.~\ref{sec:weissconductivity} 
and discuss their implications. 

{\it (i) Semiclassical  regime:} 
We address disordered graphene samples characterized by an electronic elastic mean free path $\ell$. 
The electronic transport is considered as diffusive, with sample sizes $L\gg \ell$, and semiclassical, 
with $k_F\ell\gg 1$. 
Under these assumptions, the conductivity can be predicted with good accuracy by 
Eq.~\eqref{eq:conductivity}.
For good quality graphene samples, where $\ell\geq 100$ nm, $k_F\ell\gg1$ demands typical carrier 
concentrations $|n_e| \geq 10^{11}$cm$^{-2}$, which is easy to attain in experiments.

{\it (ii) Perturbation theory:}
The evaluation of $\Delta \sigma_{yy}$ relies on using first order perturbation theory to calculate 
the semiclassical electron velocities $v_{\zeta,y}$,  Eq.~\eqref{eq:v_semiclassic}. Hence, it 
requires the Landau level spacing to be much larger than the energy correction due to the 
modulated perturbation potential.
The Landau levels that contribute to the conductivity are those close to the Fermi energy,
 corresponding to 
\begin{equation}
\label{Fermifilter}
n_F\approx
\frac{1}{2}
\left(\frac{E_F}{\hbar\omega_0}\right)^2=
\frac{1}{2}
\left(k_Fl_B\right)^2.
\end{equation}
The applicability of the perturbation theory demands that the LL spacing $E_{n+1}-E_{n}$ is large as 
compared with the correction $\Delta E_n$ given by  Eq.~\eqref{perturbation}. 

Let us consider the contributions due to strain and parallel magnetic field separately. 
For the scalar potential, $E_{n+1}-E_{n}>\Delta E_n$ constraints the Landau levels index $n$ to
\begin{equation}
\label{perturbationscalar}
n<\frac{2}{\pi^4} \left(\frac{\lambda}{l_B}\right)^2 
\left(\frac{\hbar v_F}{g\lambda} \right)^4 
\left(\frac{\lambda}{h_0}\right)^8,
\end{equation}
where we assume $n\gg1$. Using Eq.~\eqref{Fermifilter}, the above relation can be 
conveniently cast as
\begin{equation}
\label{perturbationscalar2}
k_Fl_B<
\frac{2}{\pi^2} \left(\frac{\lambda}{l_B}\right) 
\left(\frac{\hbar v_F}{g\lambda} \right)^2 
\left(\frac{\lambda}{h_0}\right)^4.
\end{equation}
The profile height parameters that govern the potential modulation enter the expression mainly 
as a $\lambda/h_0$ ratio, but the remaining quantities appear in a convoluted manner. We note 
that $k_F=\sqrt{\pi |n_e|}$ gives some freedom to easily fulfill the inequality by tuning the doping.

For the case of modulated magnetic fields,  $E_{n+1}-E_{n}>\Delta E_n$ restricts $n$ to
\begin{equation}
n<\left[\left(\frac{\hbar \pi}{2 A e \lambda} \right)^4 \left(\frac{8\pi^4 l^2_B}{\lambda^2} \right)\right]^{1/3},
\end{equation}
where $A$ = $A_0$ corresponds to the intrinsic pseudo magnetic case, while $A = A_{\|}$ stands for 
the external parallel magnetic field one. Using Eq.~\eqref{Fermifilter} we write 
\begin{equation}
\label{ }
\frac{\pi}{4}
\left(\frac{\beta_{\rm G}\kappa l_B}
{2a} \right)^{4/3}
|n_e|<\left[
\left(\frac{\lambda^2}{h^8_0} \right)\right]^{1/3}.
\end{equation}
for the case of strain generated gauge field and
\begin{equation}
\frac{\pi}{4}
\left(\frac{e B_{\|}l_B}{2\hbar \pi^2} \right)^{4/3}
|n_e|<\left[
\left(\frac{1}{h^4_0\lambda^6} \right)\right]^{1/3},
\end{equation}
for the external parallel magnetic field.

\begin{figure}[htbp]
\centering \includegraphics[width=1\columnwidth]{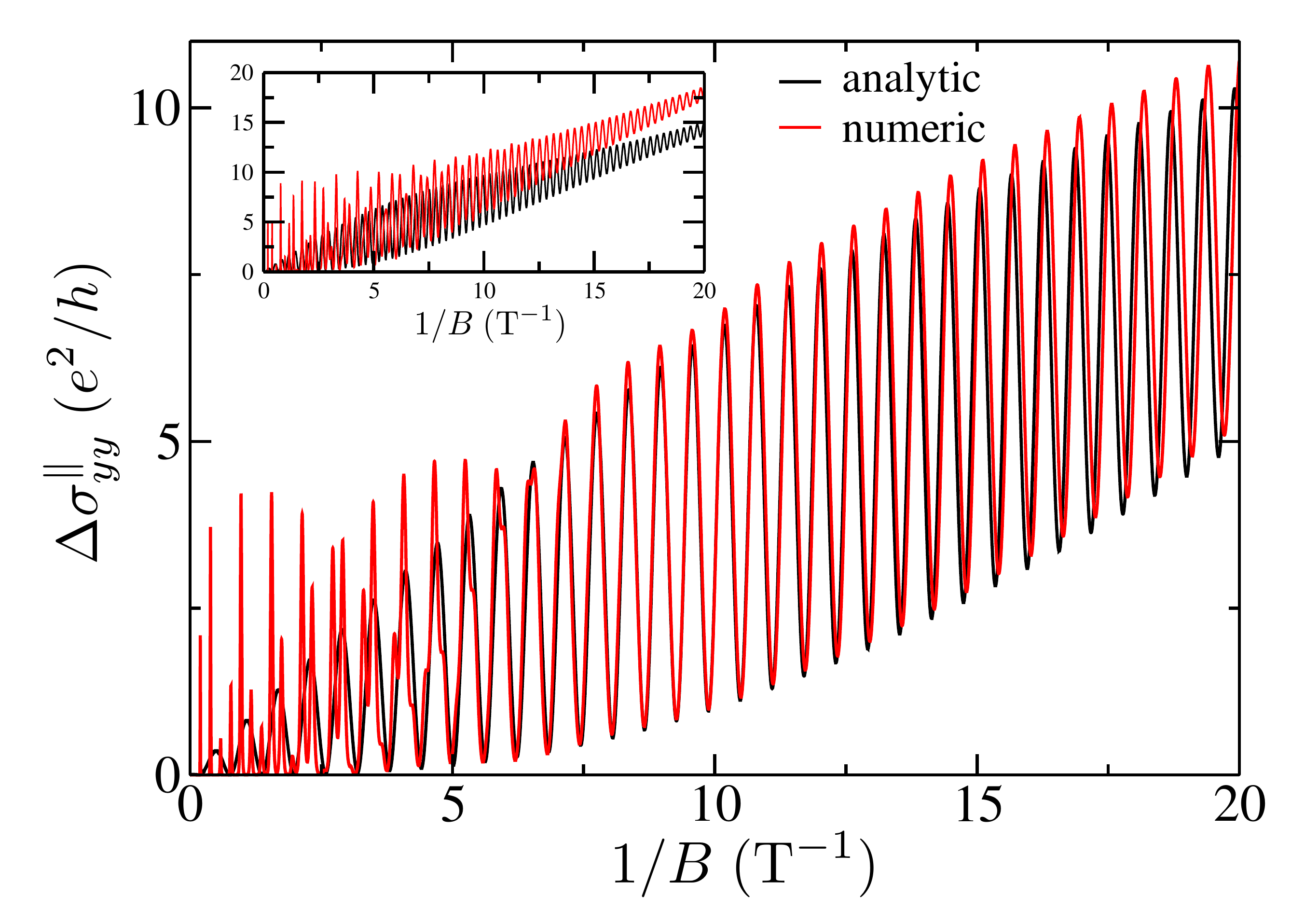}
\caption{(Color online) External magnetic field contribution to the transversal conductivity 
$\Delta \sigma^{||}_{yy}$ as a function of $1/B$ for an electronic density of $n_e=5\times 10^{11}$ cm$^{-2}$, 
$\lambda=100$ nm, $h_0=1$ nm, $B_{\|}=8$ T, $T=4$ K, and $\tau = 10^{-13}$s. Inset: Same parameters, 
except $\lambda=50$ nm.}
\label{fig:numeric_vs_analytic}
\end{figure}

{\it (iii) Asymptotic limit:} 
In Sec.~\ref{sec:weissconductivity}, Eq.~\eqref{asymptotic} is used to obtain a closed analytical 
expression for the conductivity oscillations. This asymptotic expression for the Laguerre polynomials 
requires that $n\gg1$ and $\lambda/\lambda_F\geq4$. 
In Fig.~\ref{fig:numeric_vs_analytic} we compare the ``analytic" $\Delta \sigma_{yy}$, calculated using 
Eq.~\eqref{Functionmagneticmodulation}, with the ``numeric" $\Delta \sigma_{yy}$ obtained from the 
numerical calculation of Eq.~\eqref{magneticcontributionnumeric} for a representative set of parameters.
We observe that by decreasing the magnitude of the perpendicular magnetic field, corresponding to
increasing $n$, the agreement between the analytical and the numerical results 
progressively improves, as expected. For small magnetic fields the agreement depends on 
$\lambda/\lambda_F$, as explained Sec.~\ref{scalarfield}. The inset shows 
$\Delta \sigma^{||}_{yy}$ versus $1/B$ for a case where $\lambda/\lambda_F \approx 1$. The Weiss 
oscillations persist, but their period show a small deviation from our analytical results and
the slope displays a more pronounced difference.  

\begin{figure}[htbp]
\centering \includegraphics[width=1\columnwidth]{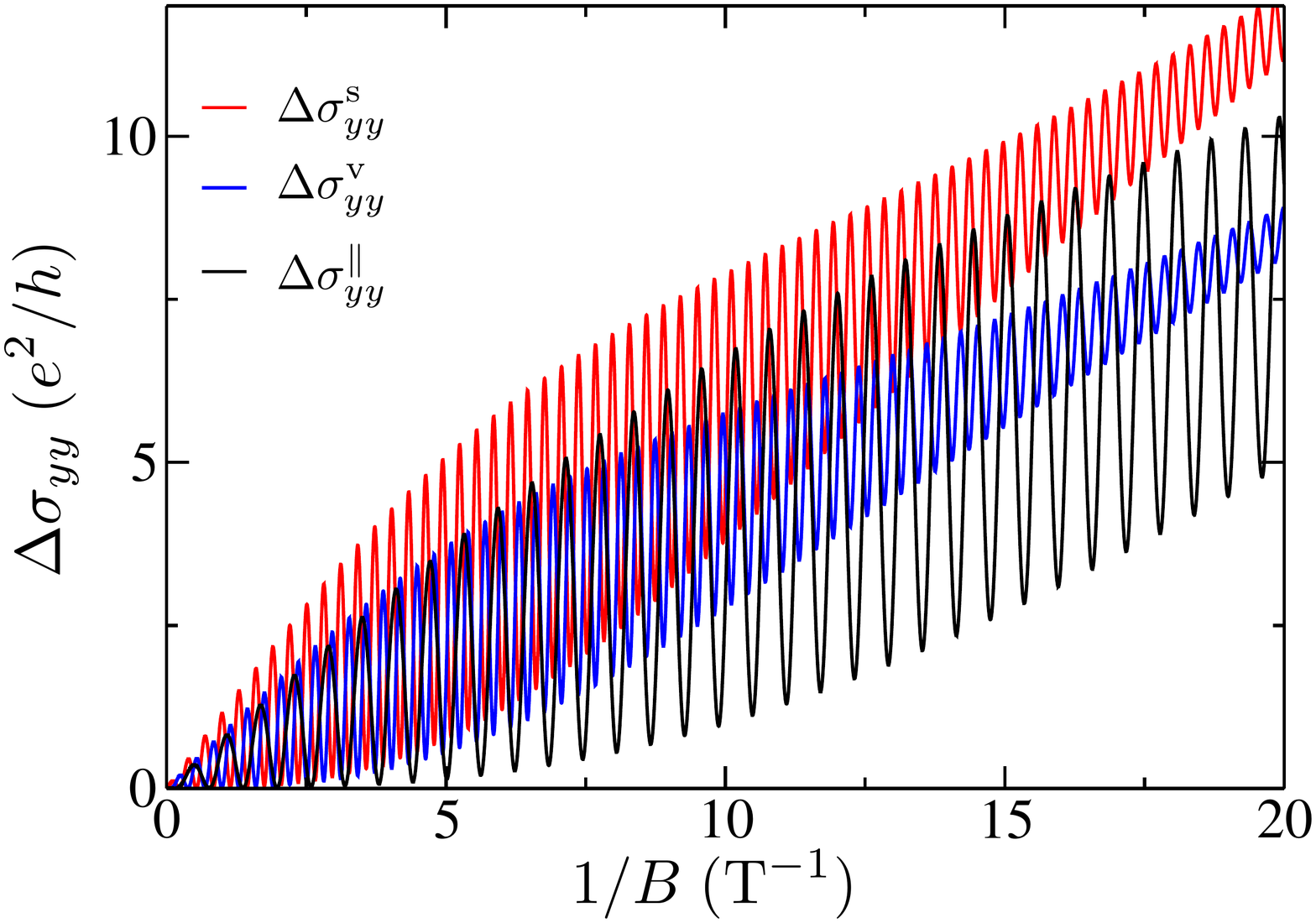}
\caption{(Color online) Magnetoconductivity $\Delta \sigma^{||}_{yy}$ as a 
function of $1/B$ for the three modulated perturbation potentials discussed in the text. 
Here we use $B_{\|}=8$T, $\lambda= 100$nm, $h_0=1$nm, $T=4$K, and $\tau\approx 10^{-13}$s.}
\label{fig:conductivity_3sources}
\end{figure}

To study the combined effect of the three modulated potentials considered in this paper, 
we use Eq.~\eqref{eq:conductivity} with the total velocity
\begin{equation}
\label{totalvelocity}
v^{\rm T}_{\zeta, y}=v^{\rm s}_{\zeta, y}+v^{\rm v}_{\zeta, y}+v^{\|}_{\zeta, y}.
\end{equation}
Following the same steps as before, we write the conductivity as a sum of three independent contributions 
\begin{equation}
\Delta \sigma_{yy}=
\Delta \sigma^{\rm s}_{yy} + \Delta \sigma^{\rm v}_{yy} + \Delta \sigma^{\|}_{yy},
\end{equation}
since upon integration over $k_y$ the cross terms average to zero. We show the three contributions 
separately in Fig.~\ref{fig:conductivity_3sources}.
For the sake of definition, $\Delta \sigma^{v}_{yy}$ represents an average over all possible lattice 
orientations. For a given experimental realization it is possible to measure the angle $\theta$.

Figure~\ref{fig:conductivity_3sources} indicates that, for the chosen set of parameters, all considered
mechanisms  contribute with similar weights to the Weiss oscillations. 
We caution that our strain calculations represent an upper limit, since we neglect atomic in-plane 
relaxations and screening. Hence, we expect the in-plane magnetic field to be the most efficient
way to study the effect. On the other hand, in view of the large quantitative uncertainty on the degrees of 
screening and in-plane relaxation in actual systems, the investigation of Weiss oscillations in the 
absence of $B_{||}$ has the potential to provide interesting insight on this issue.

In a realistic situation, a good description of the height profile certainly requires considering more 
than a single-harmonic. In such case the different contributions to the conductivity oscillations no 
longer decouple. 
Notwithstanding, it is still easy to single-out the external parallel magnetic field oscillations by 
varying $B_{||}$. For moderate values of $B_{||}$ we expect this contribution to dominate over 
pseudo-fields generated by strain. 

\begin{figure}[htbp]
\centering \includegraphics[width=1\columnwidth]{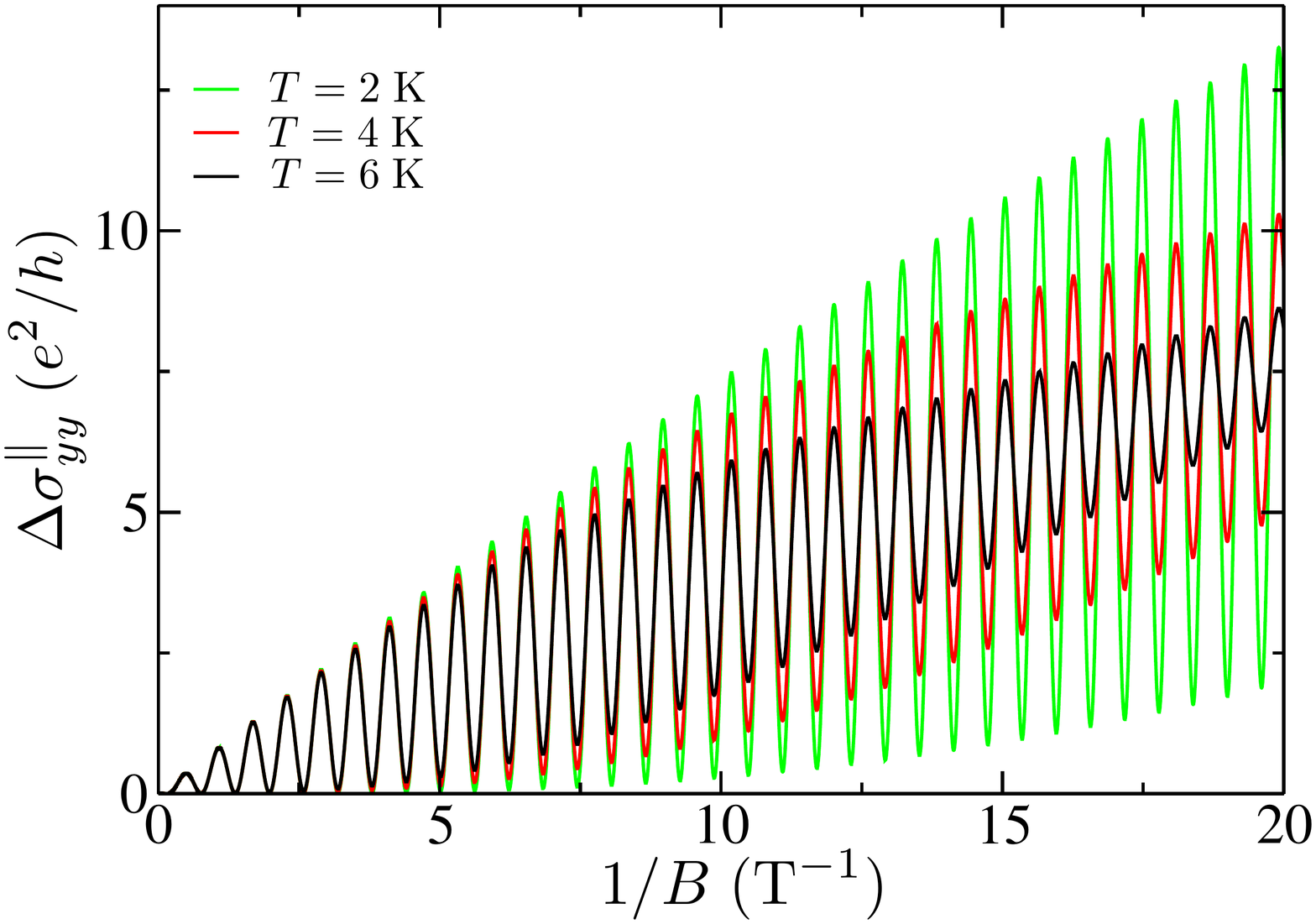}
\caption{(Color online) Conductivity as a function of $1/B$ for the external in-plane magnetic field for
different representative temperatures.
Here we take $B_{\|}=8$T, $\lambda= 100$nm, $h_0=1$nm, and $\tau\approx 10^{-13}$s.}
\label{fig:conductivity_temperature}
\end{figure}

Let us now discuss the temperature dependence. Figure~\ref{fig:conductivity_temperature}
shows the effect of the damping term $S(T/T_W)$ on the oscillation amplitude of $\Delta \sigma_{yy}$ 
for few representative temperatures. Note that $T_W\propto B$. Hence, for a fixed $T$, by decreasing 
$B$, one decreases $T_W$ and progressively quenches the Weiss oscillations (See Fig.~\ref{fig:conductivity_temperature}).

Shubnikov de Haas (SdH) oscillations are also periodic in $1/B$. Since their periodicity does not depend
on the profile heigh geometry, in principle, they are easy to distinguish from Weiss oscillations. Their 
characteristic temperatures are also very distinct: The temperature damping of the Weiss 
oscillations are given by Eq.~\eqref{damp} and 
the SdH characteristic temperature is \cite{Tan2011,Monteverde2010} $k_B T_{\rm SdH} = (\hbar \omega_0/2\pi^2)
(k_F l_B) ^{-1}$. Hence
\begin{equation}
\label{eq:TW-over-TSdH}
\frac{T_{\rm W}}{T_{\rm SdH}}= \frac{k_F \lambda}{4}.
\end{equation}
For the parameters we use $k_F \lambda \gg 1$, $T_{\rm W}/T_{\rm SdH}>1$. Thus, in general for a given
temperature we expect the SdH oscillations to be more damped than the Weiss ones.

\section{Conclusions}
\label{sec:conclusions}

We have studied the effects of a periodic profile height modulation on the electronic magnetotransport 
properties of graphene monolayer sheets. We have shown that such set up is suited for the study of Weiss 
oscillations either caused by strain fields or by an external magnetic field parallel to the graphene-substrate 
interface.

The Drude conductivity is obtained using first order perturbation theory within the effective low-energy 
Dirac Hamiltonian to calculate the semiclassical electronic velocity in the presence of a modulated 
potential. We built our analysis on the analytical results of Refs.~\onlinecite{Matulis2007,Tahir2008}. 
We consider the cases of strain induced pseudo magnetic field and pseudo electric potential, 
as well as the case of a modulated effective magnetic field originated by an external $B$-field applied parallel 
to the graphene sheet.
We studied the Weiss oscillations in the transverse conductivity in all these cases, discussing their geometry 
and temperature dependence.
By casting the expressions for the conductivity in terms of the most relevant length scales of the problem 
we were able to verify that the classical interpretation of the Weiss oscillation based on the commensurability 
of the cyclotron orbit with the modulation period \cite{Beenakker1989} still holds for Dirac-like Hamiltonian
systems, a connection that has been so far overlooked in the graphene literature 
\cite{Tahir2008,Matulis2007,Tahir2007,Tahir2007PRB}.

We presented a careful discussion of the validity range of our theory taking into account realistic experimental 
values for the carrier concentration, modulation height profile and temperature. We stablished clear distinctions 
criteria between Weiss and Shubnikov de Haas oscillations, based on the behavior of the conductivity oscillations 
with doping, substrate height profile and temperature.
Using these elements, we proposed a setup to experimentally investigate Weiss oscillations in graphene systems.

Such study can be particularly useful to provide further insight on effect of strain fields in the electronic properties of graphene, a subject of intense theoretical investigation, but still with limited quantitative experimental results.

\acknowledgments 
This work has been supported by the Brazilian funding agencies CAPES, CNPq, and FAPERJ.

\bibliography{Weiss_oscillation}

\end{document}